\documentstyle[12pt,epsfig]{article} 
\newlength{\dinwidth}                       
\newlength{\dinmargin}                      
\newcommand{\pmb}[1]{%
        \setbox0=\hbox{#1}%
        \kern-.02em\copy0\kern-\wd0
        \kern+.04em\copy0\kern-\wd0
        \kern-.02em\raise.0217em\box0}

\newcommand{\lsim}{
 \mathrel{\setbox0=\hbox{$<$}\raise0.6ex\copy0\kern-\wd0
 \lower0.65ex\hbox{$\sim$}}}
\newcommand{\gsim}{
 \mathrel{\setbox0=\hbox{$>$}\raise0.6ex\copy0\kern-\wd0
 \lower0.65ex\hbox{$\sim$}}}

\begin{document}
\leftline{\hfill OSU-96-0829}
\vspace{0.2cm}
\centerline{\Large\bf  Color Transparency and Color Opacity in Coherent}
\centerline{\Large \bf 
 Production of Vector Mesons off Light Nuclei at small x}
\vspace{0.2cm}
\centerline{\large L.~Frankfurt$^{a,d}$, V.~Guzey$^b$, 
W.~Koepf$^c$,  M.~Sargsian$^{a,e}$, M.~Strikman$^{b,d}$}
\vspace{-0.5cm}
\begin{quotation}
\noindent $^{(a)}$ Tel Aviv University, Tel Aviv, Israel \\
\noindent $^{(b)}$ Pennsylvania State University, University Park, PA, USA \\
\noindent $^{(c)}$ Ohio State University, Columbus, OH, USA \\
\noindent $^{(d)}$ S.Petersburg Nuclear Physics Institute, Russia \\ 
\noindent $^{(e)}$ Yerevan Physics Institute, Yerevan, Armenia  
\end{quotation}
\vspace{-1.0cm}
\begin{quotation}
\noindent
{\bf Abstract:}
Coherent electroproduction of vector mesons off $^2$H and $^4$He 
is considered in the  kinematics of  deep inelastic scattering. 
Special emphasis is given to the 
$-t\sim 0.8-1$ GeV$^2$ region
where the cross section is dominated by  the interaction  of the
$q\bar q$ configuration in the $\gamma^*$ with two nucleons. These 
kinematics provide the unique possibilities to
study quantitatively in vector meson production 
the onset of Color Transparency 
with increasing $Q^2$
as well as its gradual disappearance at very small $x$, i.e.,
perturbative Color Opacity.
\end{quotation}

\vspace{-0.4cm}

\noindent{\Large\bf 1 \  Introduction}

Recent theoretical analyses \cite{BFMGS,AFS,FKS} 
have demonstrated that, in the limit of large $Q^2$, exclusive production of
vector mesons in the process $\gamma_L^*+p\rightarrow V+p$ is controlled by 
the calculable interplay of hard perturbative QCD and soft nonperturbative 
physics. 
Recent experimental data \cite{HERA}  seem to indicate that the hard
mechanism starts to dominate already at $Q^2 \ge 5 $ GeV$^2$.

In the small $x$ region, vector meson production
in the target rest frame 
is essentially a three stage process. First, at 
a distance $l_c \sim {1 \over 2 m_Nx}$ before the target, the
$\gamma^*_L$ transforms into
a small transverse size $q\bar q$ pair where $b_{q\bar q} \equiv 
r_{qt}- r_{\bar q t} \approx 3/Q$, i.e., $b_{q\bar q}(Q^2 \sim 10$ GeV$^2) 
\approx 0.4$ fm.
Then, the small $q\bar q$  pair  interacts with the target with an amplitude
\cite{BBFS}:
\vspace{-0.15cm}
\begin{equation}
A(q\bar q T)\mid_{t=0} = {Q^2\pi^2\over 3x} 
\left[ b^2 \alpha_s(Q^2)\cdot \left(i - {\pi\over 2}{d\over lnx}\right)
xG_T(x,Q^2)\right],
\label{ef5}
\vspace{-0.15cm}
\end{equation}
where $G_T(x,Q^2)$ is the target's gluon density.
The third stage, the transformation  $q \bar q \rightarrow V$ occurs 
long after the target at distances $\sim 2q_0/m_V^2$.
The use of completeness over the diffractively produced states allows to 
express the result in terms of bare parton distributions 
within the target and the vector meson similar to 
DIS processes \cite{BFMGS}:
$A^{\gamma^*T \rightarrow VT} = 
\psi^{\gamma^*\rightarrow q\bar q} \otimes A(q\bar q T) 
\otimes \psi^{q\bar q\rightarrow V}$,
where $\psi^{\gamma^*\rightarrow q\bar q}$ is the 
wave function of the $\gamma^*\rightarrow q\bar q$ transition,  
$\psi^{q\bar q \rightarrow V}$ is the $q \bar q$ component of the 
$V$'s wave function. $A(q\bar q T)$ describes the scattering of the $q\bar q$ 
off the target $T$ with a cross section given by  Eq.(\ref{ef5}). 

The use of nuclei allows to check the QCD prediction that
the interaction cross section of a small $q\bar q$ with a nucleon 
is indeed small, i.e., Color Transparency,
and whether it can reach (due to the increase of the gluon density at small
$x$) values comparable to those for the interaction of light hadrons (pions),
i.e., perturbative Color Opacity. One possible strategy is to study coherent
vector meson production off heavy enough nuclei at $t \sim 0$.
Another possibility, which we consider here, is to use
the process $\gamma^* + A \rightarrow V + A$ with the lightest nuclei (A=2,4)
for $V=\rho, \rho', \omega, \phi...$ at $x \le 10^{-2}$, and 
to focus on the region
of comparatively large $|t|$, where {\bf the
interaction with two nucleons dominates.}
We restrict ourselves to the coherent channel because this 
case can be singled out
experimentally in an unambiguous way. There is clear experimental
evidence for  the dominance of the rescattering diagrams  in
the coherent production off the deuteron 
at $|t|\ge 0.6$ GeV$^2$ (see Ref.\cite{dsh} and references therein). 
Due to the quadrupole contribution, the diffractive minimum is filled up. 
This is the  reason why our next choice is a $^4He$ target, which is  
spherical and without quadrupole form factors, and  for which the
diffractive minimum  occurs  at $-t \sim 0.2$ GeV$^2$. There, as a result,
one is sensitive to the rescatterings at considerably smaller $|t|$.

\medskip

\noindent{\Large\bf 2 \  Cross Section}

At small $x$ and large $Q^2$,  where the
average  $ b_{q\bar q}$  is small, nuclear effects are  generated through 
double rescattering of the $q\bar q$  pair off the target nucleons. 
As a result of QCD evolution, the parton wave function of a compact 
configuration can evolve to normal transverse hadronic size.
Such configurations will lead to shadowing effects in the leading
power of $Q^2$. This effect is not too small  at $t=0$ 
at sufficiently small
$x$ \cite{AFS}. Thus, in QCD, there is no direct relation between the
smallness of the cross section and the smallness of secondary interactions. 
A strict QCD prediction is that double scattering effects should
decrease with  increasing $Q^2$ more fastly than the single
scattering amplitudes \cite{BFMGS}. But an actual calculation of the
double scattering term is model dependent at the moment.  
Hence, we account for double scattering of the $q\bar q$ pair, which is 
numerically large, and neglect leading twist effects due to  QCD 
evolution of the $q\bar q$ pair to normal hadronic size,
which decrease with $t$ more rapidly.
For the  deuteron, we obtain (suppressing spin indices):
\vspace{-0.15cm}
\begin{equation}
{d\sigma^{\gamma^*_L{^2}H\rightarrow V^2H}\over dt}  =  
{1\over 16\pi}\left|2f^{(1)}(t)S_d({q\over 2}) + 
{i\over 2}f^{(2)}\left({t\over 2}\right)
\int {d^2 k'_\perp\over (2\pi)^2}S_d(k'_\perp)e^{-Bq^{`2}}\right|^2
\label{ef7}
\vspace{-0.15cm}
\end{equation}
where $t\approx -q^2_\perp$ and 
$S_d(k) = F_c(q) + \left(3{(Jq)^2\over q^2}-1\right)F_Q(q)/\sqrt{2}$.
$F_C$ and $F_Q$ represent the deuteron's charge and quadrupole form factors. 
Similarly, for coherent scattering off $^4He$, we obtain:
\vspace{-0.15cm}
\begin{equation}
{d\sigma^{\gamma^*_L{^4}He\rightarrow V^4He}\over dt} = {1\over 16\pi} \left| 
4f^{(1)}(t)\Phi(t) + 
{i3f^{(2)}({t\over 2})\over  4\pi(\alpha+B)} e^{{\alpha\over 8}t}
\right|^2
\label{ef12}
\vspace{-0.15cm}
\end{equation}
where $\Phi(t)$ is the charge form factor of $^4He$ 
($\Phi(t) \approx \exp(3\alpha t/3)$ for small $t$).
We restrict ourselves here to the double scattering amplitude since
multi-scattering amplitudes violate energy-momentum conservation due to the
production of multi-particle states. So, their contribution
should be zero (S.Mandelstam cancelation) 
within the approximation when only a $q\bar q$ pair  is
considered. 
In Eqs.(\ref{ef7}) and (\ref{ef12}), the multiple scattering
amplitudes are defined as:
\vspace{-0.15cm}
\begin{equation}
 f^{(n)} \sim 
 \int d^2 b \, \psi_{\gamma^*}^L(b) \psi_V(b) \sigma^n_{q \bar q N}(b) 
 exp(Bt/n)
 \label{ef13}
\vspace{-0.15cm}
\end{equation}
where $\sigma_{q\bar q}$ is given by Eq.(\ref{ef5}). Because of the 
small size of $q\bar q$, the slope  of the 
elementary amplitude,   $B\approx 2.5$ GeV$^{-2}$ \cite{HERA},
is mainly determined by  the nucleon's two-gluon  form factor.

\medskip

\noindent{\Large\bf 3 \ Onset of Color Transparency (CT)}

The onset of CT leads to a rather nontrivial dependence of the 
coherent production cross section on $x$,  $t$ and $Q^2$. To estimate 
those effects, we analyze the ratio $R(t) = {d\sigma^{\gamma_L^*}
\over dt}(t)/{d\sigma^{\gamma_L^*}\over dt}(t=0).$
It follows from Eqs.(\ref{ef7}-\ref{ef13}) that at $-t\ge -t_0\sim
0.5$ GeV$^2$ ($> 0.2$ GeV$^2$ for $A=4$),
when the coherent cross sections are dominated by the rescattering terms, 
$R(|t|>|t_0|)\sim x^2G_N^2(x,Q^2)e^{Bt}/Q^4$.
Therefore, one should expect a strong decrease of $R$ with
increasing $Q^2$. At the same time, for fixed $Q^2$, one expects a 
fast increase of the cross section with decreasing  $x$. This 
increase is restricted by a unitarity condition \cite{FKS}. 
An observation of  a slowing down of such an increase at small 
$x$ would be a clear signature of the onset of color opacity. 
For $Q^2 \sim 5$ GeV$^2$, such a effect could occur already at HERA energies.

 In Fig.1 we present  $R(t)$ for fixed $Q^2$ and different $x$, calculated 
for coherent scattering off $^2H$ and $^4He$.
We show also in this figure the expectation of the Vector Dominance model
(VMD)  in which a $\rho$-meson is produced in the first interaction 
and then scatters off the nucleons with a cross section similar to the $\pi N$ 
cross section. In the case of a $^4He$ target, one expects a clean 
minimum which is very sensitive to the amount of rescattering 
in Eq.(\ref{ef13}). The
investigation of the depth of the diffraction minima would allow to 
check another prediction of QCD, namely the large value of the real 
part of the production amplitude, $ReF/ImF\sim 0.3-0.4$. 

\begin{figure}[t]
\vspace{-1.1cm}
\hspace{-8.cm}  
\centering{
\epsfig{figure=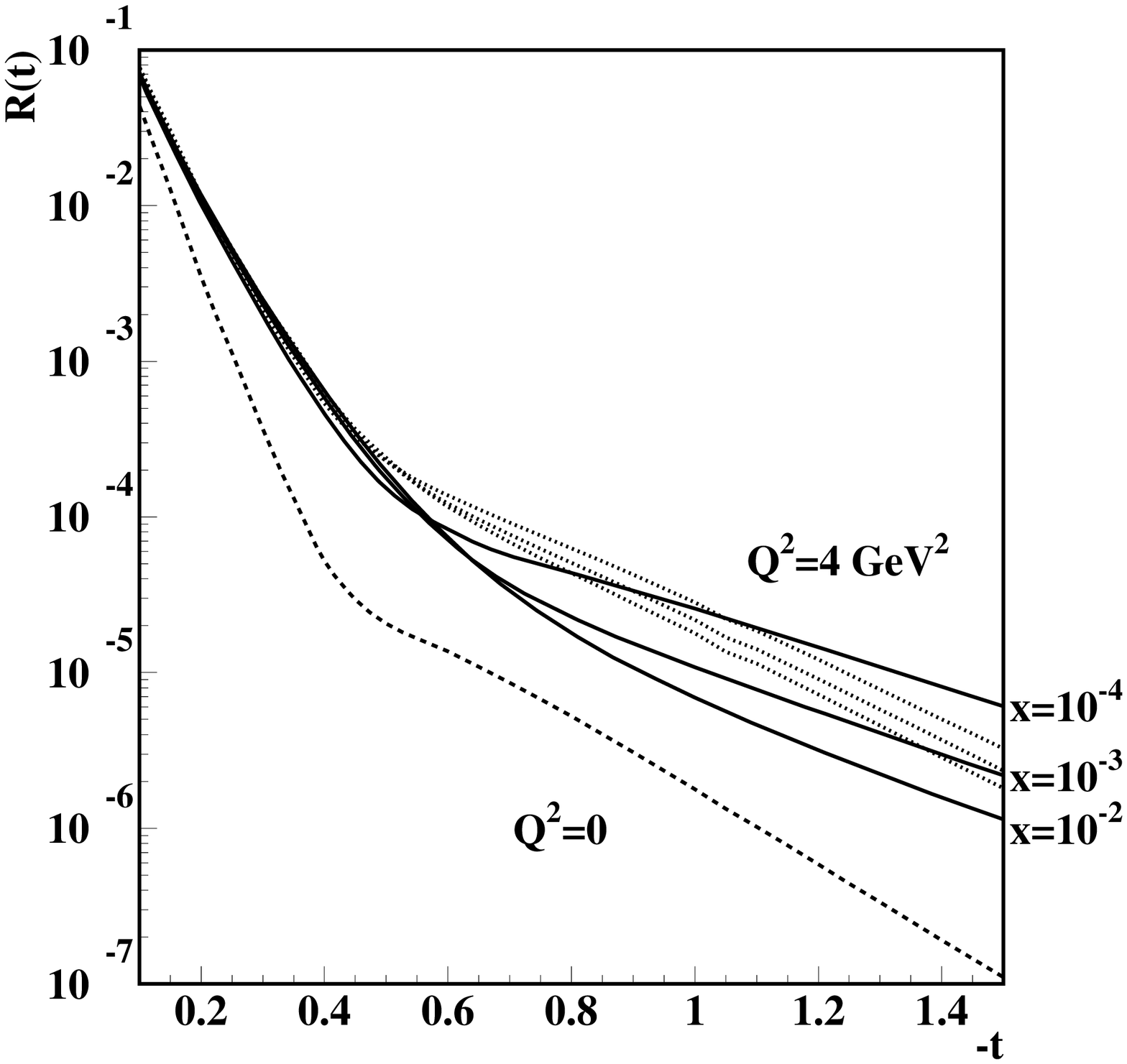,height=7.6cm}}
\end{figure}
\begin{figure}
\vspace{-7.95cm}
\hspace{8.cm}  
\centering{
\epsfig{figure=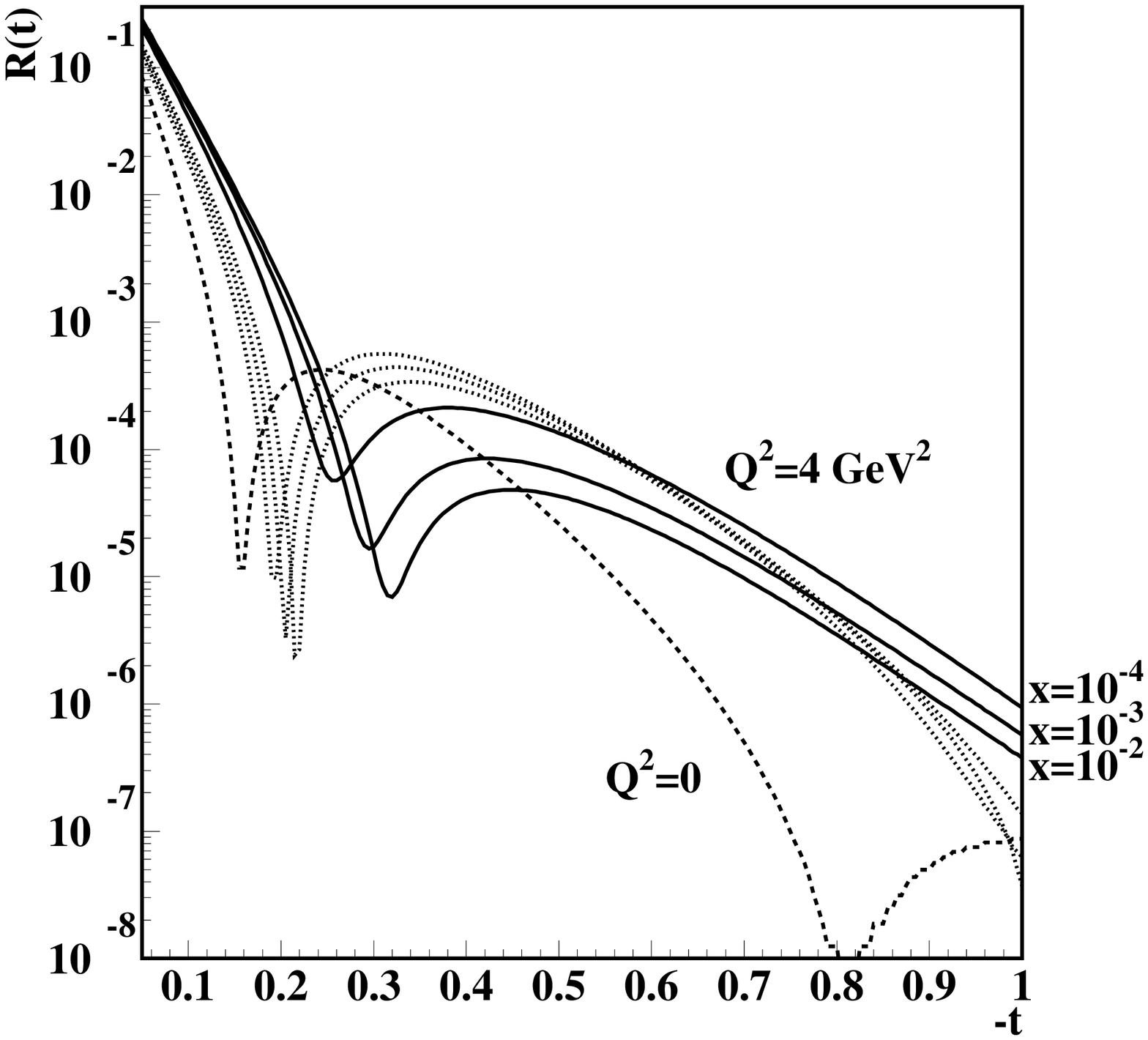,height=7.6cm}}

\vspace{-0.5cm}

\caption {\it $R(t)$ as a function of $t$ at different 
$x$. Solid curves are QCD calculations as explained in text, dotted  
curves are VMD predictions, and the dashed curve is  for $Q^2=0$
(which for $^4$He includes $n>2$ rescattering terms).}

\end{figure}

\medskip

\noindent{\Large\bf 3 Critical Assessments and Conclusions}

In this discussion we assumed that the  only mechanism that generates 
nuclear effects is  double  rescattering, and we neglected 
the leading twist mechanism of multiple scattering related 
to leading twist nuclear shadowing. It may compete with 
the mechanism we discussed above in a certain $x$ and $Q^2$ range. 
This question requires further studies, and a detailed experimental 
study of the $t$, $x$ and $Q^2$ dependencies in a large kinematical 
range is necessary.
However, an important signature, which will persist in a wide
kinematical range, is that the secondary rescatterings would result 
in a strongly different shape of the $t$ dependence as compared to 
the case of a real photon projectile. 
If  such a difference will disappear at
very small $x$ close to the unitarity bound, this would establish the 
$x$ and $Q^2$ range for the onset of the Color Opacity phenomenon.


\vspace{-0.8cm}

\baselineskip=8pt

\begin{thebibliography}{96}

\vspace{-0.25cm}

\bibitem{BFMGS} S.~J.~Brodsky et al,
 Phys.~Rev. {\bf D50}, 3134 (1994).
\vspace{-0.25cm}

\bibitem{AFS} H.~Abramowicz, L.~L.~Frankfurt and M.~I.~Strikman
              DESY-95047; SLAC Summer Inst. 1994: 539-574.
\vspace{-0.25cm}

\bibitem{FKS} L.~L.~Frankfurt, W.~Koepf and M.~I.~Strikman,
               Phys. Rev. {\bf D54}, 3194 (1996). 
\vspace{-0.25cm}

\bibitem{HERA}ZEUS, M.~Derrick et al., Phys. Lett. {\bf B356}, 601 (1995);
Contribution to XXVII Int. Conf. on High Energy Physics, Warsaw, July 1996.
\vspace{-0.25cm}

\bibitem{BBFS} B.~Blattel, G.~Baym, L.L.~Frankfurt and 
               M.~I.~Strikman Phys. Rev. Lett. {\bf 70}, 896 (1993).
\vspace{-0.25cm}

\bibitem{dsh}T.~H.~Bauer, R.~D.~Spital, D.~R.~Yennie and F.~M.~Pipkin,
 Rev. Mod. Phys. {\bf 50}, 261 (1978).

\end{thebibliography}
\end{document}